\documentclass{PoS}

\usepackage{multicol}
\usepackage{multirow}
\usepackage{graphicx}
\usepackage{comment}

\newcommand{\veritas}{{VERITAS}}
\newcommand{\magic}{{MAGIC}}

\newcommand{\swift}{\textit{Swift}}
\newcommand{\xrt}{\textit{Swift}-XRT}

\newcommand{\psr}{PSR J2032+4127}
\newcommand{\TeVJ}{TeV J2032+4130}
\newcommand{\mt}{MT91 213}

\newcommand{\GR}{$\gamma$-ray}
\newcommand{\xray}{X-ray}
\newcommand{\msol}{M$_{\odot}$}
\newcommand{\degree}{$^\circ$}
\newcommand{\code}[1]{\texttt{#1}}
\title{\xray\ and TeV \GR\ emission from the 50-year period binary system \psr/\mt}

\ShortTitle{TeV \GR s from \psr/\mt}

\author{\speaker{Tyler Williamson}, for the \veritas\thanks{https://veritas.sao.arizona.edu/} and \magic\thanks{https://magic.mpp.mpg.de} collaborations\footnote{for collaboration lists see PoS(ICRC2019)1177.}\\
        University of Delaware\\
        E-mail: \email{tjwilli@udel.edu}}

\abstract{
We report on X-ray and TeV \GR\ observations of the pulsar/Be star binary \psr/\mt. \psr\ is a 143-ms \GR\ pulsar which shares a long period (45-50 year) and highly eccentric orbit with the massive Be star \mt. TeV \GR\ emission was detected from the binary following a coordinated observing campaign over the fall 2017 periastron with \veritas, \magic, and \xray\ monitoring with \xrt. The discovery of this \GR\ binary makes it just the second such source known to contain a pulsar as the compact object. We report on over 100 hours of extensive TeV observations across the periastron passage, which reveal variations in the TeV flux by an order of magnitude over time scales of days. The \xray\ flux was also found to be highly variable, although it was not directly correlated with the TeV flux. These observations present serious challenges to existing models of the system, which will require significant revisions. We also discuss the steady and extended TeV source \TeVJ, which lies in the same direction as the binary system, and its potential association with the pulsar.
}

\FullConference{36th International Cosmic Ray Conference -ICRC2019-\\
		July 24th - August 1st, 2019\\
		Madison, WI, U.S.A.}

\begin{document}

\section{Introduction}
\GR\ binaries comprise a massive star orbiting together with a compact companion with a peak in $\nu F_{\nu} > 1$ MeV \cite{DUBUS}. With a population of only eight systems \cite{TeVCat}, they represent a relatively small source class. The emission in these systems is thought to be powered by either the interaction of a pulsar wind with the stellar wind / disk of the massive companion, or by relativistic jets powered by the accretion of matter from the massive companion onto the compact object (a neutron star or black hole). \GR\ binaries can thus act as laboratories for studying particle acceleration in a continually and periodically changing physical environment. 

\psr\ is a \GR\ \cite{LAT_PSR} and radio \cite{RADIO_PSR} pulsar located in the Cygnus OB2 region. It was recently identified as the compact object orbiting with the $\sim$ 15 \msol\ B0Ve star \mt\ \cite{Lyne} in a highly eccentric orbit with a period of 45-50 years \cite{Ho}. \psr\ lies at the edge of the extended very high energy (VHE; E > 100 GeV) \GR\ source \TeVJ\ \cite{HEGRA_TeV2032}. \TeVJ\ was the first unassociated VHE source, and, although a pulsar wind nebula association with \psr\ is favored \cite{VTS_TeV2032}, its origins remain unknown. The discovery of the binary nature of \psr\ raises new questions about the relationship between \TeVJ\ and \psr.

The binary system \psr/\mt\ reached periastron on 13 November 2017. The system was well studied in the period leading up to and covering periastron. Rapidly increasing X-ray flux was the first evidence of the interaction between the winds of \psr\ and \mt\ \cite{Ho}. This increasing \xray\ flux was later found to be variable on time scales of weeks \cite{Li_17}, likely a consequence of clumps in the stellar wind of \mt. Multi-wavelength observations of the system across periastron were first reported in \cite{Li_18}, finding strong variablility in the \xray\ flux, but only steady emission in the GeV flux (possibly due to the dominance of the pulsar magnetosphereic interaction in the GeV band). Optical and \xray\ spectra were used in \cite{Coe_19} to model the geometry of the circumstellar disk of \mt\ with respect to the orbit of the binary. Here we summarize the detection of a VHE source consistent with \psr/\mt\ by the \veritas\ and \magic\ collaborations, first reported in \cite{VTS_binary}, and compare observations with models of the system and with other \GR\ binaries.

\section{Observations and Analysis}
\subsection{\xray\ Observations}
The \xray\ telescope aboard the Neil Gehrels \swift\ Observatory (\xrt\,\cite{SwiftXRT}) was designed for detection of \GR\ bursts and is sensitive to photons between 0.2 and 10 keV. A total of 210 observations (490 ks, including 117 ks across the fall 2017 periastron passage) from the period of 2008 June 16 - 2018 December 10 were analyzed. The observations were taken in photon counting mode and processed with the HEAsoft analysis package. The data were cleaned using the standard \code{xrtpipeline} tool. Each observation was then background subtracted using several regions offset from the pointing position. Spectra were extracted from a 20 pixel radius region around \psr\, and then fit to an absorbed power law using the \code{xspec} \cite{xspec} spectral fitting software. The entire data set is well fit to an absorbed power law model with $\Gamma = 1.76 \pm 0.04$ and $N_H = (0.91 \pm 0.03)\times 10^{22}\ \mathrm{cm}^{-2}$.

\subsection{VHE Observations}
\veritas\ \cite{veritas} is an array of four 12-m Imaging Atmospheric Cherenkov Telescopes (IACTs) located in Southern Arizona. The instrument has a 3.5\degree\ field of view and is sensitive to \GR s from  $\sim$ 85 GeV to 30 TeV. \veritas\ and its performance are further described in \cite{VERPerf}.

\magic\ \cite{magic} consists of two 17-m IACTs located in La Palma, Spain. The array has a 3.5\degree\ field of view and has a sensitive energy range of $\sim$ 30 GeV to 30 TeV. \magic\ and its performance are further described in \cite{MAGPerf}. 

\begin{figure}
    \centering
    \includegraphics[width=0.7\textwidth]{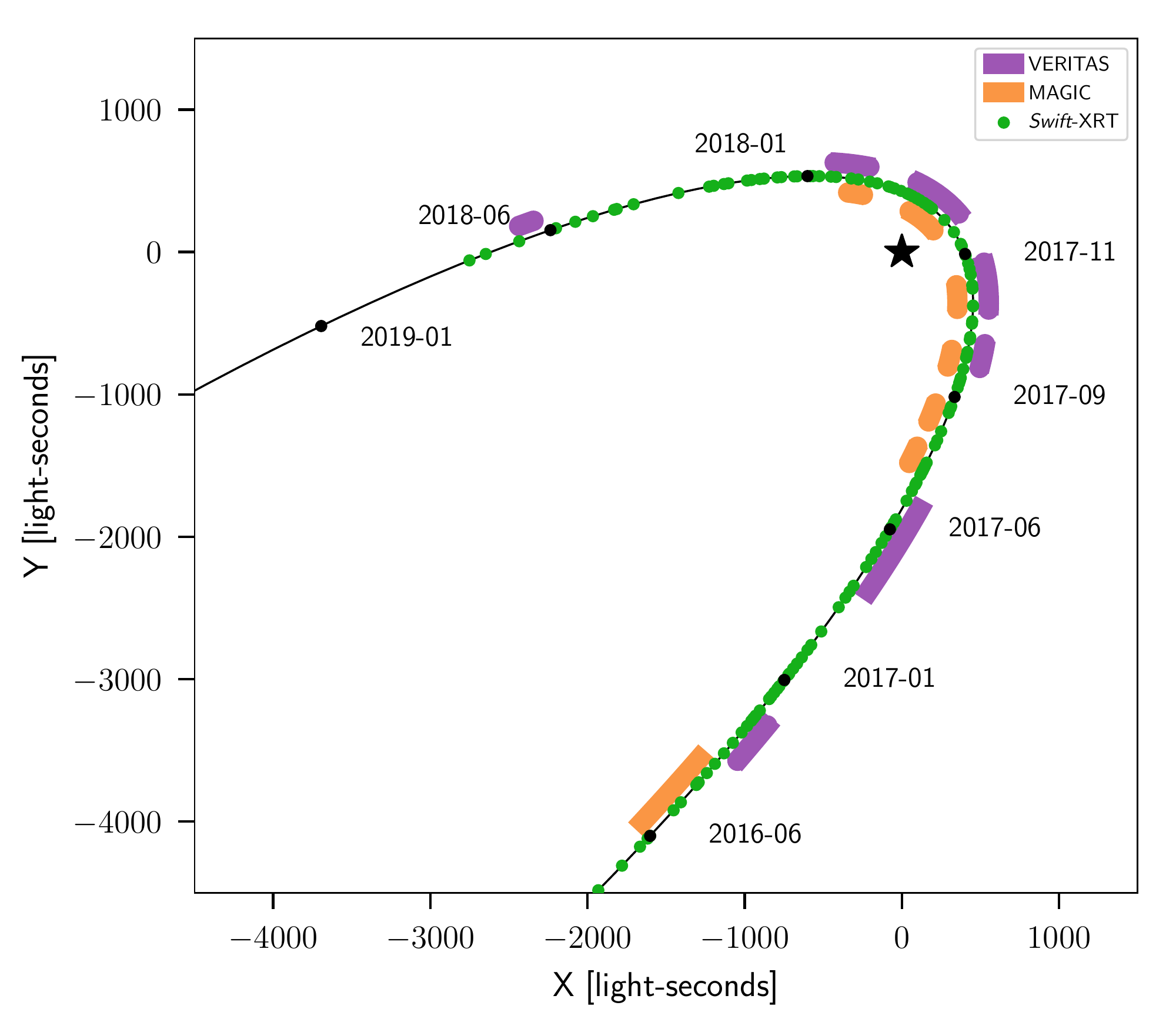}
    \caption{Observation times around the binary orbit for \xrt,\ \veritas,\ and \magic,\ using orbital model 2 from \cite{Ho}.}
    \label{fig:obs}
\end{figure}

Between 2016 September and 2018 June, \veritas\ observed \psr\ for a total of 142.3 hours, including 99.6 hours covering the periastron passage between 2017 September and December. \magic\ observed the source for 87.9 hours, including 34.2 hours between 2017 June and December. Approximately 52 hours of \veritas\ archival data, taken prior to 2016, were also re-analyzed in order to study the extended emission from \TeVJ\ before the appearance of the binary interaction. Orbital coverage of periastron for all observations are shown in Figure \ref{fig:obs}.

Observations for both instruments were conducted in the standard ``wobble'' mode \cite{fomin94}, keeping the source offset from the camera center to allow for background estimation within the field of view. Data were analyzed using the tools described in (\cite{VTS_ana}, \veritas, \cite{MAG_ana}, \magic).

\section{Results}
\begin{figure}[!htb]
\centering
\begin{tabular}{cc}
  \includegraphics[width=65mm]{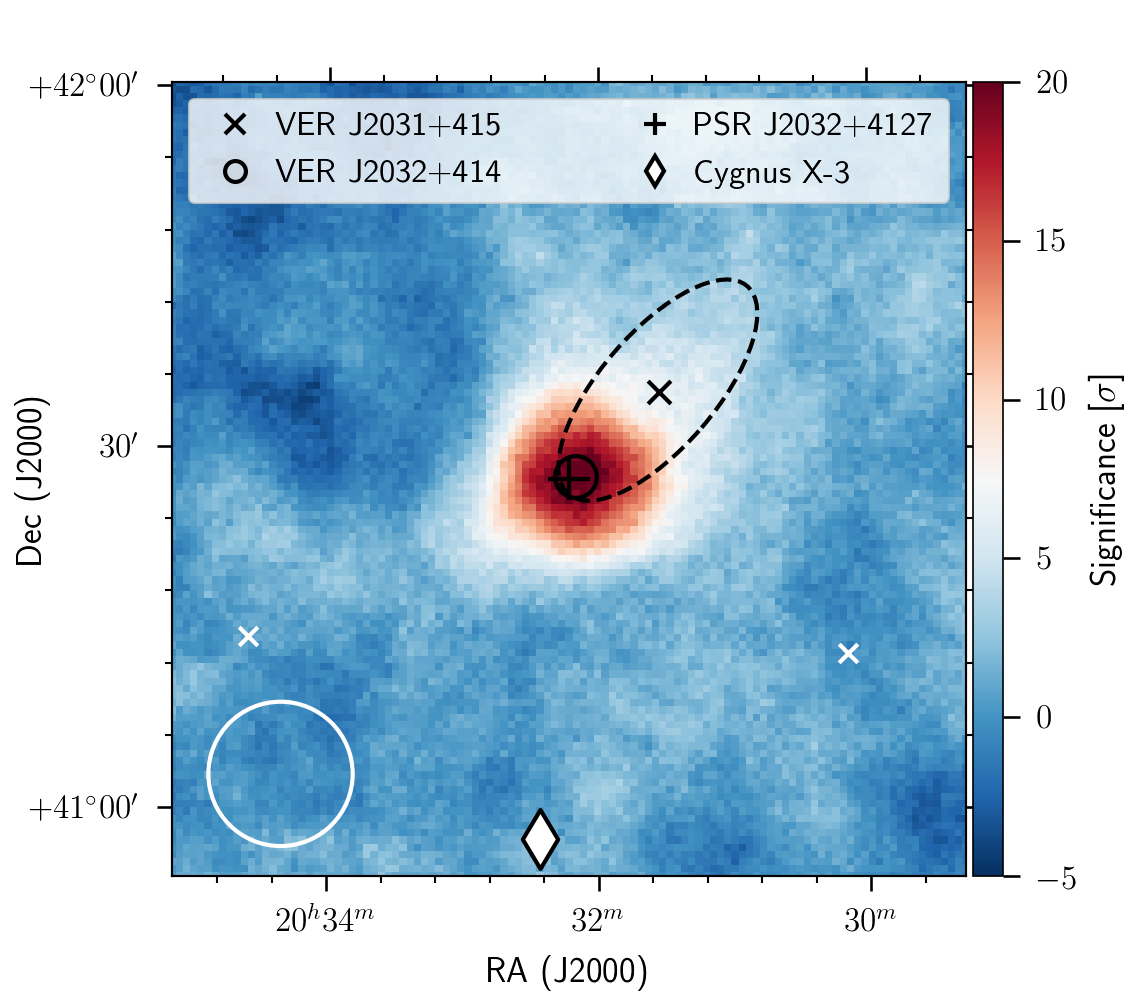} &   \includegraphics[width=65mm]{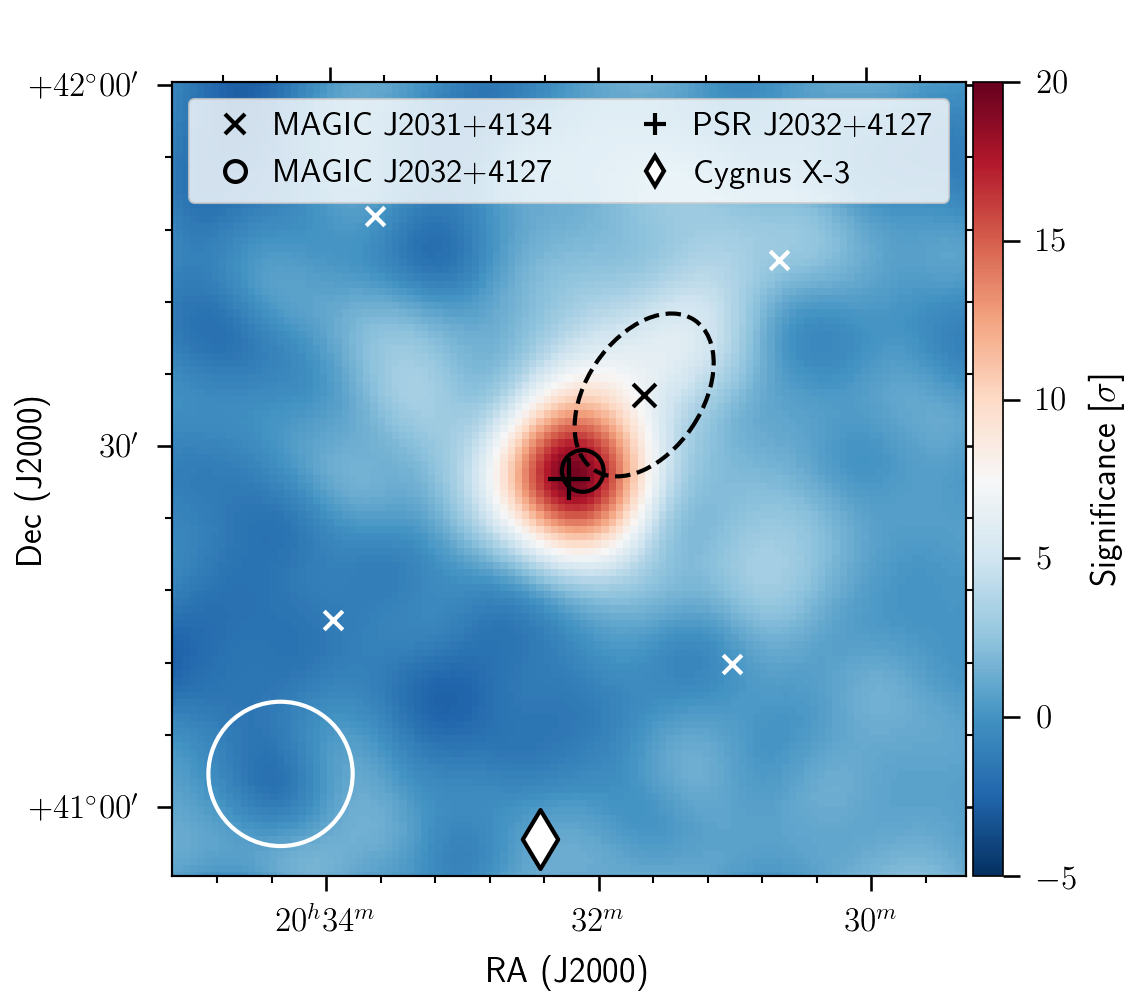} \\
(a) \veritas\ & (b) \magic \\
\end{tabular}
\caption{Significance sky maps for each instrument. The \GR\ point spread function of the instrument is shown as a white circle in the lower left corner. The morphology of \TeVJ,\ as measured by each instrument, is drawn as an ellipse which overlaps with the centroid of emission from the binary system.}
\label{fig:skymap}
\end{figure}

The VHE observations covering the periastron passage in the fall of 2017 resulted in a significant detection of a \GR\ source coincident with the position in the sky of \psr/\mt. \veritas\ and \magic\ both strongly detect the source, with significances of 21.5 standard deviations ($\sigma$), and 19.5 $\sigma$, respectively. Significance sky maps for both instruments are shown in Figure \ref{fig:skymap}. 

%Skymaps
%\begin{figure}[!htb]
%\centering
%\begin{subfigure}{.5\textwidth}
%  \centering
%  \includegraphics[width=\linewidth]{fig2a.png}
%  \caption{A subfigure}
%  \label{fig:skymapa}
%\end{subfigure}%
%\begin{subfigure}{.5\textwidth}
%  \centering
%  \includegraphics[width=\linewidth]{fig2b.png}
%  \caption{A subfigure}
%  \label{fig:skymapb}
%\end{subfigure}
%\caption{A figure with two subfigures}
%\label{fig:skymap}
%\end{figure}

\begin{figure}[!htb]
\centering
\begin{tabular}{cc}
  \includegraphics[width=65mm]{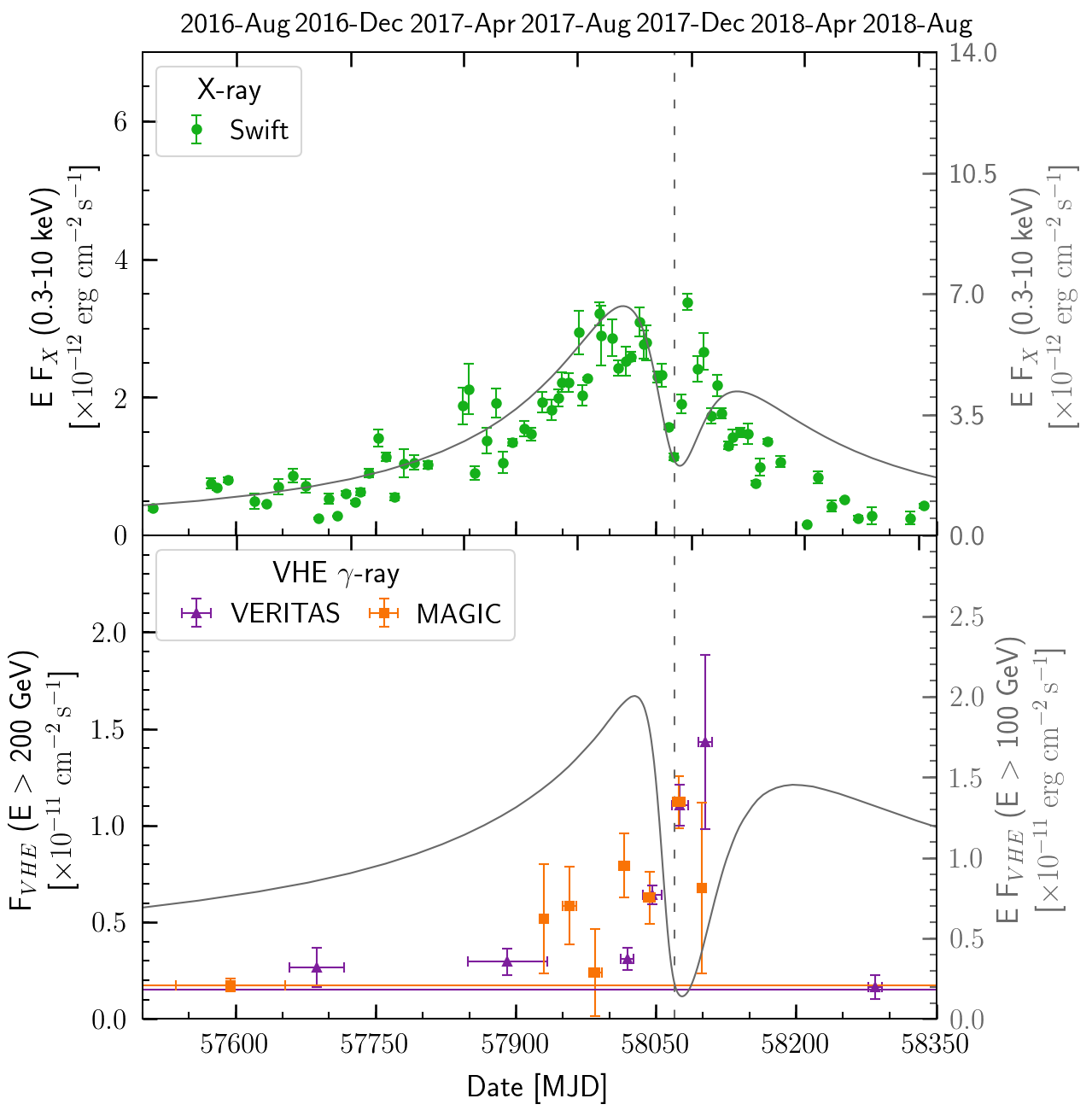} &   \includegraphics[width=65mm]{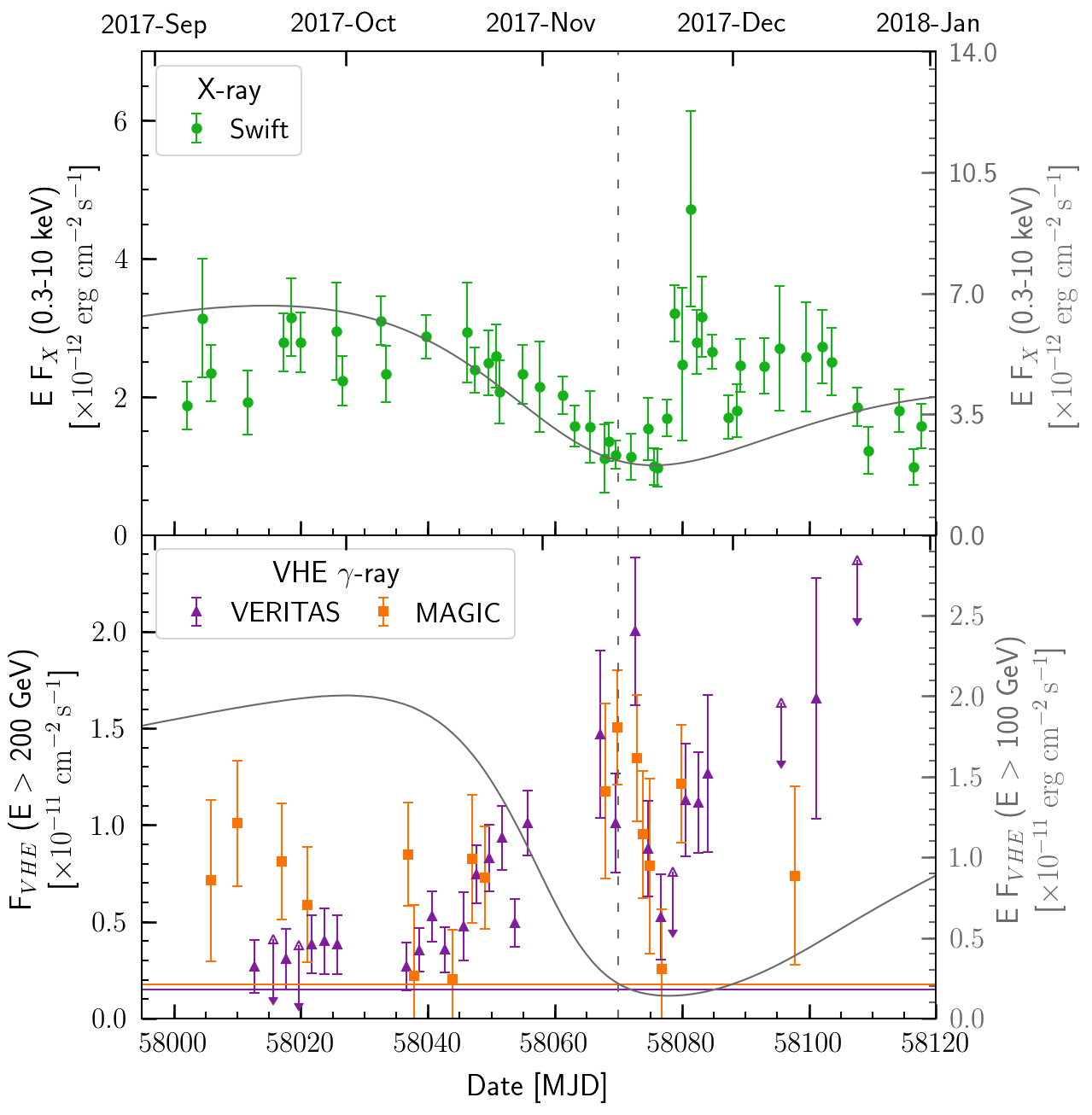} \\
(a) Long-term light curve& (b) Light curve across periastron\\
\end{tabular}
\caption{\xray\ (top panels) and VHE light curves. Figure (a) shows the long-term light curve starting in early 2016, with week-long bins for the \xray\ flux. Figure (b) shows the light curve with finer bins and zoomed to show detailed variability around periastron. Shown in gray, and corresponding to the scale on the right-hand side, are the predicted \xray\ and VHE light curves from \cite{takata} and \cite{Li_18}. Periastron is indicated by the vertical dashed line.}
\label{fig:LC}
\end{figure}

Light curves for both \xray\ and VHE observations are shown in Figure \ref{fig:LC}, along with a model light curve generated by \cite{takata} and \cite{Li_18}. Both light curves show a general increasing trend in flux leading up to periastron followed by a dip shortly after, although the time scales and exact times for these features differ across the wavelengths and in general the \xray\ and VHE emission are not in phase. Both light curves are highly variable over the periastron passage, in some cases showing variations in flux of nearly an order of magnitude on time scales as short as a few days. The minimum for both \xray\ and VHE emission is attained at or shortly after periastron.

%lightcurve
%\begin{figure}[!htb]
%\centering
%\begin{subfigure}{.5\textwidth}
%  \centering
%  \includegraphics[width=\linewidth]{fig1a.png}
%  \caption{A subfigure}
%  \label{fig:skymapa}
%\end{subfigure}%
%\begin{subfigure}{.5\textwidth}
%  \centering
%  \includegraphics[width=\linewidth]{fig1b.png}
%  \caption{A subfigure}
%  \label{fig:skymapb}
%\end{subfigure}
%\caption{!Placeholder figure!}
%\label{fig:LC}
%\end{figure}

\subsection{Spectral Results}\label{sec:spectra}
The energy spectrum of the binary interaction was determined by modeling the emission from the region as a superposition of flux from the binary interaction on top of the steady baseline emission from the nearby extended source \TeVJ.\ The emission from \TeVJ\ at the location of \psr\ is modeled as a power law, with spectral index constrained by the 1 $\sigma$ range in \cite{VTS_TeV2032}. For the binary interaction, a power law with and without an exponential cutoff was tested.

\begin{figure}[!htb]
\centering
\begin{tabular}{cc}
  \includegraphics[width=65mm]{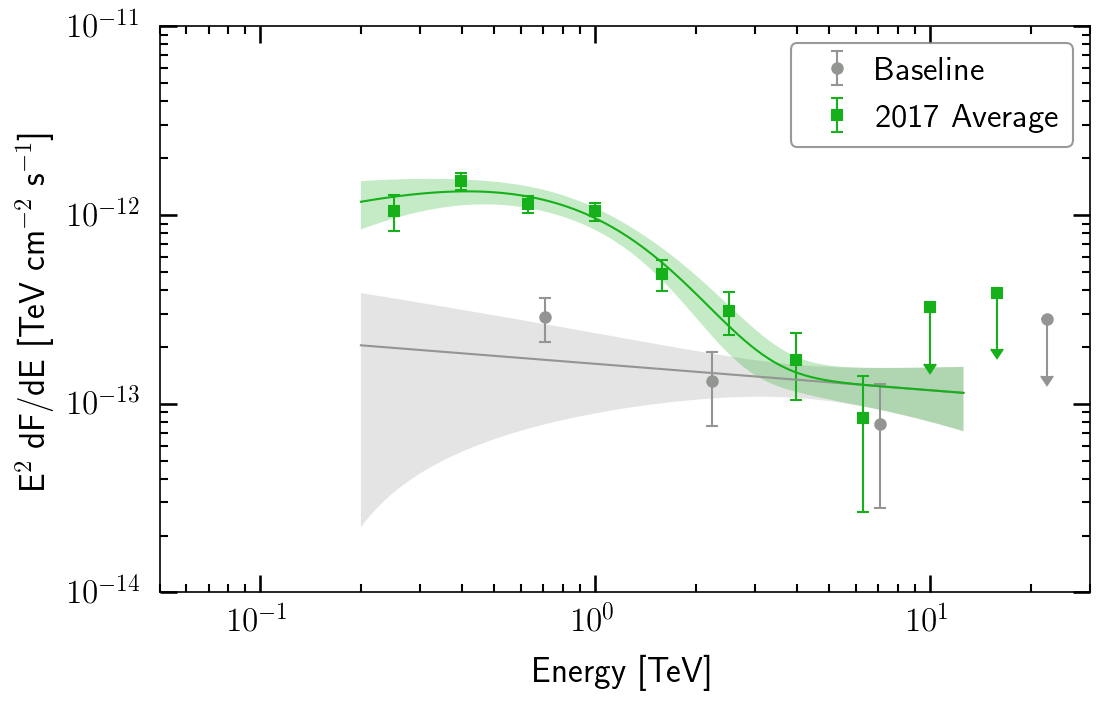} &   \includegraphics[width=65mm]{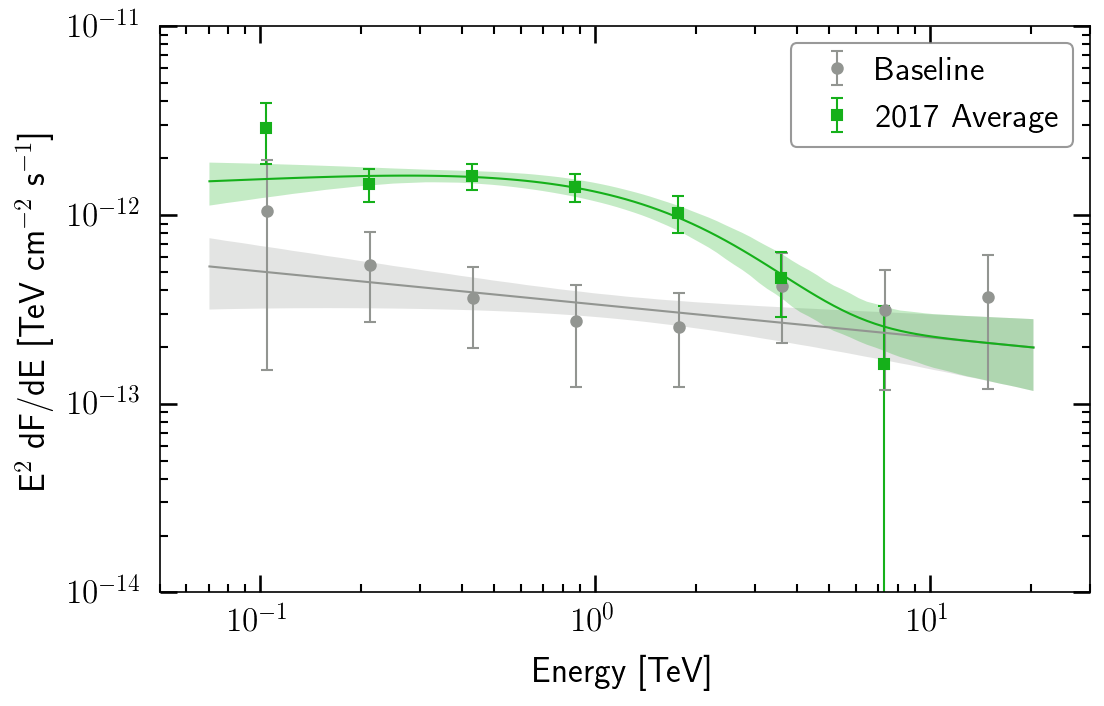} \\
(a) \veritas\ 2017 Average & (b) \magic\ 2017 Average \\
 \includegraphics[width=65mm]{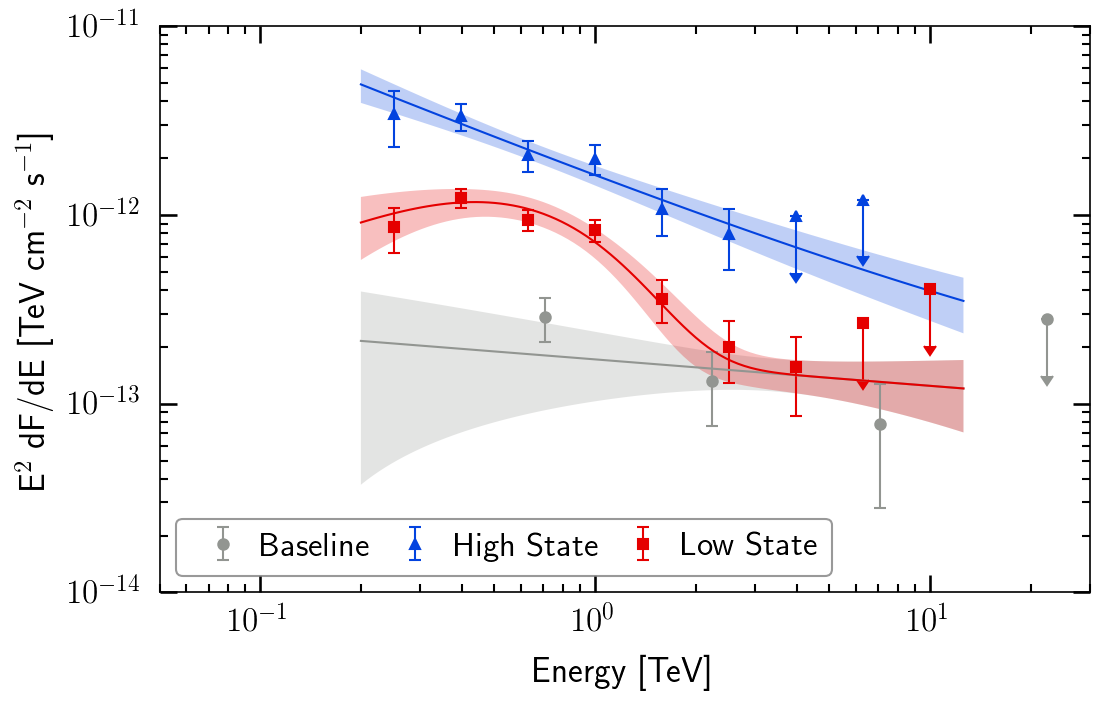} &   \includegraphics[width=65mm]{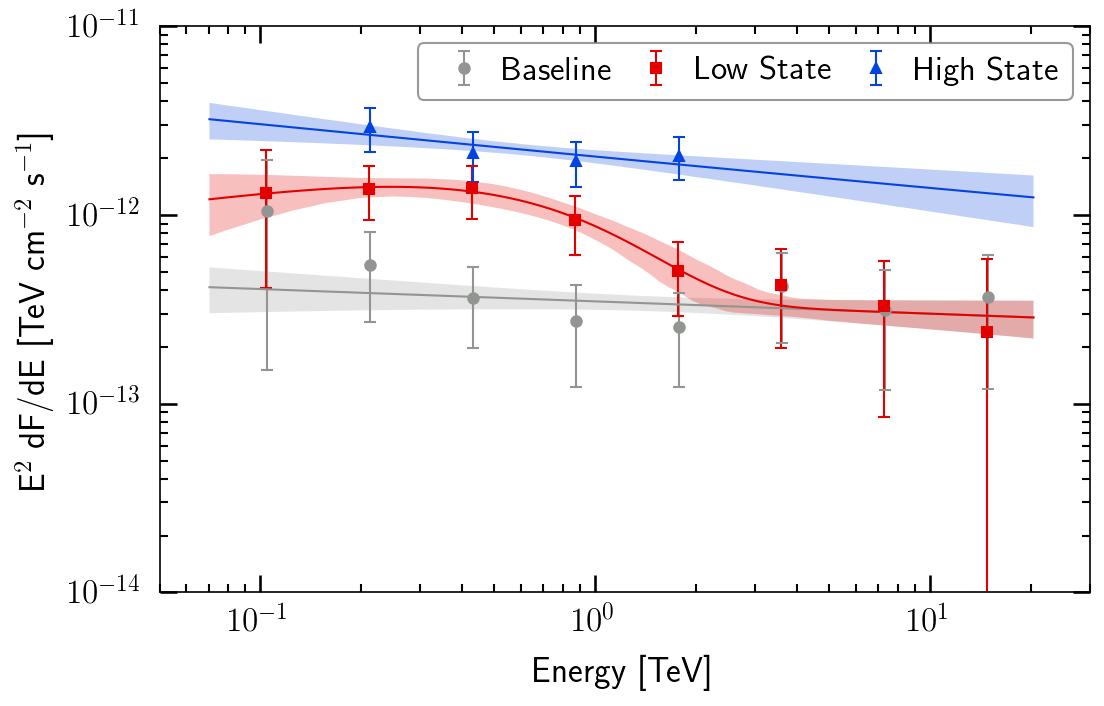} \\
(c) \veritas\ High and Low states & (d) \magic\ High and Low states \\
\end{tabular}
\caption{Spectral energy distributions and fits for the entire 2017 data set, and split by flux state. The butterflies show the 1 $\sigma$ statistical uncertainty of the fit.}
\label{fig:spectra}
\end{figure}

The power-law component for the baseline emission was simultaneously used to model the steady emission from the region in the re-analyzed archival data, prior to the appearance of the binary interaction. Both models (baseline plus binary, and baseline alone) were jointly fit; the results of the fit are shown in Figure \ref{fig:spectra} and Table \ref{tab:spectra}. For the binary component, both instruments significantly favor a power law with a low-energy exponential cutoff.

Spectra were also extracted after binning the data according to flux state, to search for spectral variability. The above fit was performed again after splitting the 2017 data into two periods, defined as the ``high'' state, where the flux above 200 GeV exceeds $1.0\times 10^{-11}\mathrm{cm}^{-2}\mathrm{s}^{-1}$ and the ``low'' state, which covers the remaining data, prior to periastron. The results of this fit are shown in Figure \ref{fig:spectra}. The low-state spectrum shows a clear cutoff for both instruments in the 300 - 600 GeV range. For \veritas\, the high state is equally well fit with and without a cutoff, and so the simpler model is assumed. In the case of \magic\, insufficient statistics in the high state prevented testing a power law with an exponential cutoff model.

%spectra
%\begin{figure}[!htb]
%\centering
%\begin{subfigure}{.5\textwidth}
%  \centering
%  \includegraphics[width=\linewidth]{fig3a.png}
%  \caption{A subfigure}
%  %\label{fig:skymapa}
%\end{subfigure}%
%\begin{subfigure}{.5\textwidth}
%  \centering
%  \includegraphics[width=\linewidth]{fig3b.png}
%  \caption{A subfigure}
%  %\label{fig:skymapb}
%\end{subfigure}
%\caption{!Placeholder figure!}
%\label{fig:spectra}
%\end{figure}

\begin{center}
    
\begin{table}
\small 
\centering
\begin{tabular}{lccccccc}
\hline 
%Header
 &
Period &
Model &
N$_0$ &
E$_0$&
$\Gamma$&
E$_\mathrm{C}$&
$\chi ^2$/dof\\
 &
 &
 &
[cm$^{-2}$ s$^{-1}$ TeV$^{-1}$] &
[TeV] &
 &
[TeV]&
\\
\hline
\hline
\multirow{5}*{\centering \veritas}
& Fall 2017 & PLEC & (8.04 $\pm$ 3.37) $\times 10^{-12}$ & 0.64 & 1.26 $\pm$ 0.45 & 0.57 $\pm$ 0.20 & 8.6/6 \\

\cline{2-8}

& Low State & PLEC & (1.63 $\pm$ 1.12) $\times 10^{-11}$ & 0.56 & 0.65 $\pm$ 0.75 & 0.33 $\pm$ 0.13 & \multirow{2}*{7.9/9} \\
& High State & PL & (1.45 $\pm$ 0.18) $\times 10^{-12}$ & 1.00 & 2.73 $\pm$ 0.15 & - & \\

\cline{2-8}

& Low State & PLEC & (1.64 $\pm$ 1.12) $\times 10^{-11}$ & 0.56 & 0.65 $\pm$ 0.75 & 0.33 $\pm$ 0.13 & \multirow{2}*{7.2/8} \\
& High State & PLEC & (1.20 $\pm$ 0.41) $\times 10^{-11}$ & 0.51 & 2.37 $\pm$ 0.50 & 2.39 $\pm$ 3.23 &  \\

\cline{2-8}

\hline 

\multirow{3}*{\centering \magic}
& Fall 2017 & PLEC & $(3.77\pm1.68)\times 10^{-12}$&0.70&$1.74\pm0.37$&$1.40\pm0.97$&9.6/12\\
\cline{2-8}
& Low State & PLEC & $(5.11\pm3.61)\times 10^{-12}$&0.70&$1.55\pm0.61$&$0.58\pm0.33$&\multirow{2}*{3.0/14}\\
& High State & PL & $(1.65\pm0.14)\times 10^{-12}$&0.70&$2.20\pm0.40$&-& \\

\cline{2-8}

\hline 
\end{tabular}

\caption{Spectral properties for the binary component of the spectral fit described in section \ref{sec:spectra}. Properties shown are: model used power law (PL) or power law with an exponential cutoff (PLEC), power law normalization $N_0$, calculated at the decorrelation energy $E_0$, spectral index $\Gamma$, cutoff energy $E_\mathrm{C}$ where applicable, and $\chi^2$ per degree of freedom for the joint fit. Models grouped within a row were all fit simultaneously together with a baseline model for \TeVJ.}
\label{tab:spectra}
\end{table}

\end{center}

\section{Discussion and Conclusion}
\psr/\mt\ is the second \GR\ binary where the identity of the compact companion is confirmed to be a pulsar, after PSR B1259-63 \cite{psr59_2005}. In such systems, the observed non-thermal emission is thought to be produced by the collision between the stellar wind/disk and pulsar wind. This collision results in a shock which accelerates particles in the pulsar wind. Synchrotron radiation and inverse Compton scattering of photons from the companion star then produce \xray\ and TeV \GR\ emission, respectively. Variability in this emission is introduced by a number of factors, including the changing distance of the shock with respect to the pulsar, variation of the scattering angle in the inverse Compton process, Doppler boosting of the post-shocked pulsar wind, and absorption of high-energy \GR s by stellar photons \cite{DUBUS}.

\begin{figure}
    \centering
    \includegraphics[width=0.7\textwidth]{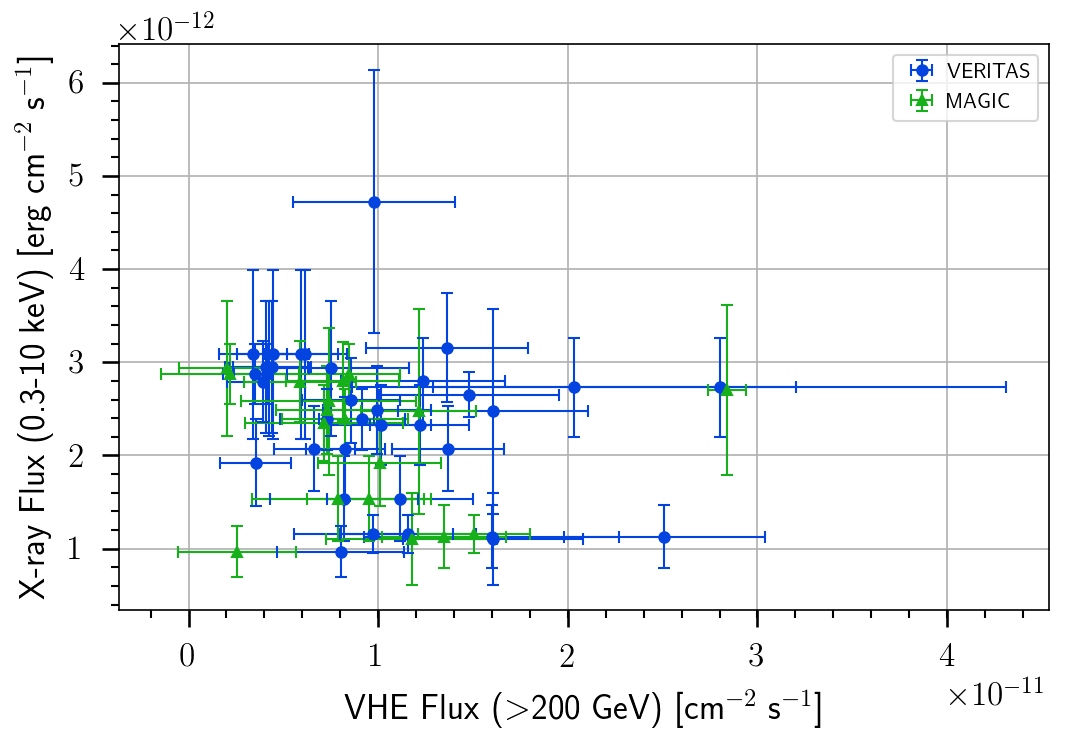}
    \caption{VHE flux vs \xray\ flux around periastron (Figure \ref{fig:LC}b).}
    \label{fig:corr}
\end{figure}

Observations of non-thermal emission from the wind collision in such systems allow for investigations into the properties of the pulsar wind, including the magnetization and momentum \cite{takata,bednarek}. The time-dependent \xray\ and VHE emission for \psr/\mt\ is modeled in \cite{takata} and revised in \cite{Li_18} with the addition of \xray\ observations during periastron. A set of light curves from this model, using a radially dependent wind magnetization, are shown in Figure \ref{fig:LC}. The modeled \xray\ emission is in general agreement with the observed \xrt\ light curve prior to periastron, but is unable to explain the rapid increase in \xray\ flux immediately following periastron. As noted in \cite{Li_18}, this could be evidence for an interaction between the pulsar and circumstellar disk of the companion star, a hypothesis potentially supported by the disappearance of radio pulsations during the same time period \cite{Coe_19}. Modeling of the VHE emission will require significant revision. A noteworthy feature of the VHE light curve is a sharp dip occurring approximately one week after periastron. The location of this feature is predicted by \cite{takata} and is similar to the post-periastron dip observed in PSR B1259-63 \cite{psr59_2005,psr59_2014}, which has been attributed to photon-photon absorption during superior conjunction of the system \cite{sushch}. While the location of this dip is accurately modeled, the feature is much more abrupt in the VHE observations than expected; indeed overall the VHE observations show variation on shorter time scales than those anticipated by the modeled emission. Another notable feature is the lack of predicted correlation between the time-dependent \xray\ and VHE emission (shown in Figure \ref{fig:corr}), which distinguishes this system from most of the other known \GR\ binaries, including PSR B1259-63. While both \xray\ and VHE emission attain their minimum at approximately the same time, other features are notably different. The \xray\ flux begins decreasing several weeks before periastron, after the inferior conjunction of the system and therefore the maximum of the Doppler beaming effect in the direction of the observer. During this same period, the VHE flux is increasing, attaining a maximum a couple of days after periastron before rapidly dropping to its minimum near the same time as the minimum of the \xray\ emission. Both fluxes briefly recover before again decaying back toward the level of baseline emission. The decay in VHE emission is less well constrained due to lack of visibility of the source for both \veritas\ and \magic\ for several months, but observations in the spring of 2018 confirm that the VHE emission from the region has returned to pre-2017 levels.

GeV emission from this region was investigated in \cite{Li_18}, which found no evidence for a new source associated with the binary interaction. It should be noted that any GeV emission originating from the intra-binary interaction could be dominated by the pulsed emission in the magnetosphere of the pulsar. A robust timing analysis should be investigated to gate out the pulsed emission and further examine evidence of GeV emission related to the binary.

The interpretation of \TeVJ\ deserves a closer look in the light of this new source. \TeVJ\ was deemed likely to be a pulsar wind nebula associated with \psr\ in \cite{VTS_TeV2032}, but the binary nature of the pulsar has since been discovered, as well as the new variable VHE source. As is briefly discussed in \cite{VTS_binary}, the luminosity and extension of \psr\ and \TeVJ\ are only marginally consistent with the population properties reported in the recent survey of pulsar wind nebulae by the HESS collaboration \cite{hesspwn}. It should be noted however that the distribution of these properties reported in \cite{hesspwn} are quite broad, and the morphological arguments laid out in \cite{VTS_TeV2032} still point to a pulsar wind nebula as the most likely explanation for \TeVJ.

\section{Acknowledgements}
\veritas\ is supported by the U.S. Department of Energy, the U.S. National Science Foundation, the Smithsonian Institution, and by NSERC in Canada. We acknowledge the excellent work of the support staff at the Fred Lawrence Whipple Observatory and at collaborating institutions in the construction and operation of VERITAS.

We acknowledge \textit{Fermi} and \textit{Swift} GI program grants 80NSSC17K0648 and 80NSSC17K0314.

The \magic\ Collaboration thanks the funding agencies and institutions listed in:

https://magic.mpp.mpg.de/ack\_201805.

\end{document}